\newcommand\mrk{Mrk 766}
\newcommand\etal{et al. }

\newcommand\xmm{{\it XMM-Newton}}
\newcommand\asca{{\it ASCA}}
\newcommand\sax{{\it BeppoSAX}}
\newcommand\nh{N_{H}}
\documentclass{aa}
\begin{document}

%   \thesaurus{02     % A&A Section 2: Physical data and processes
%              (02.01.2; % Accretion, accretion disks
%               02.02.1; % Black hole physics
%               11.19.1; % Galaxies: Seyfert
%               13.25.3)} % X-rays: general
%
   \title{The variable \xmm\ spectrum of Markarian 766
\thanks{Based on observations obtained with XMM-Newton, an ESA science 
    mission with instruments and contributions directly funded by 
    ESA Member States and the USA (NASA)}}
\authorrunning{Page \etal}

   \author{M.J. Page\inst{1},
           K.O. Mason\inst{1},
%           Th. Boller\inst{2},
           F.J. Carrera\inst{2},
           J. Clavel\inst{3},
           J.S. Kaastra\inst{4},
           E.M. Puchnarewicz\inst{1},
           M. Santos-Lleo\inst{3},
           H. Brunner\inst{5},
           C. Ferrigno\inst{4},
           I.M. George\inst{6},
           F. Paerels\inst{4},
	   K.A. Pounds\inst{7},
           S.P. Trudolyubov\inst{8}
          }

   \offprints{M.J. Page (mjp@mssl.ucl.ac.uk)}

   \institute{$^{1}$
              Mullard Space Science Laboratory, University College London,
              Holmbury St Mary, Dorking, Surrey, RH5 6NT, UK\\
%              $^{2}$
%              MPI for Extraterrestrial Physics, 85740 Garching, Germany\\
              $^{2}$
              Instituto de F\'\i sica de Cantabria (Consejo Superior de
              Investigaciones Cient\'\i ficas--Universidad de Cantabria), 
              39005 Santander, Spain\\
              $^{3}$
              XMM-Newton Science Operations Centre, 
              Astrophysics Division, ESA Space Science Department, 
              P.O. Box 50727, 28080 Madrid, Spain\\
              $^{4}$
              Space Research Organization of the Netherlands,
              Laboratory for Space Research, Sorbonnelaan 2, 
              Utrecht, CA NL-3584, Netherlands\\
              $^{5}$
              Astrophysikalisches Institut Potsdam, Sternwarte 16, 
              D-14482 Potsdam, Germany\\
              $^{6}$
              NASA Goddard Space Flight Center, Code 660, Greenbelt, 
              MD 20771, USA\\
              $^{7}$
              X-Ray Astronomy Group, Department of Physics and Astronomy, 
              Leicester University, Leicester LE1 7RH, UK\\
              $^{8}$
              Los Alamos National Laboratory, NIS-2, Los Alamos, 
              NM 87545, USA\\
%         \and
%             University of Alexandria, Department of Geography\\
%             email: c.ptolemy@hipparch.uheaven.space
%             \thanks{The university of heaven temporarily does not
%                     accept e-mails}
             }

   \date{Received ?; accepted ?}

\abstract{
The narrow-line Seyfert 1 galaxy \mrk\ was observed for 60 ks with
the \xmm\ observatory. The source shows a complex X-ray spectrum. 
The 2-10 keV spectrum can be adequately represented by a power law and broad
Fe K$\alpha$ emission.
%Clear
%evidence is found for broad Fe K$\alpha$ emission at 6.4 keV. Apart from the
%iron line, the spectrum above 2 keV can be adequately represented by a
%power-law. 
Between 0.7 and 2 keV the spectrum is harder and exhibits a flux
deficit with respect to the extrapolated medium energy slope. Below 0.7 keV,
however, there is a strong excess of emission. The RGS spectrum shows an
edge-like feature at 0.7 keV; the energy of this feature is inconsistent
with that expected for an OVII edge from a warm absorber.
\mrk\ varies by a factor of $\sim$ 2 in overall count rate in the EPIC and
RGS instruments on a timescale of a few thousand seconds, while
no significant flux changes are
observed in the ultraviolet with the OM. The X--ray variability is
spectrally dependent with the largest amplitude variability 
occurring in the 0.4-2 keV
band. The 
spectral variability can be explained by a change in flux and slope
of the medium energy continuum emission, superimposed on a less variable (or
constant) low energy emission component. 
}

%
%________________________________________________________________

\maketitle

\section{Introduction}
\label{sec:introduction}

X--ray emission currently provides the best tracer of conditions close to the
supermassive black hole central engines of AGN. However the origin of AGN
X--ray emission remains poorly understood. Between 1 and 20 keV AGN generally 
have a
power law spectrum modified by the effects of reflection and absorption (Pounds
\etal \cite{pounds90}); below 1 keV many AGN also show some `soft excess'
emission (Walter \& Fink \cite{walter93}). The majority of recent models for
AGN X--ray emission explain the soft excess as emission from 
an optically thick accretion disk, and the power law component as Compton
upscattering of soft photons by a hot corona.

\mrk\ is a nearby
(\(z=0.0129\)), soft X--ray bright (0.5 - 2 keV flux of $\sim 10^{-11}$ erg
s$^{-1}$ cm$^{-2}$) and variable (factor 2 in \(\sim\) 1000s, Leighly \etal
\cite{leighly96}) narrow line Seyfert 1 galaxy observed through a relatively
small Galactic column of \(\sim 1.8 \times 10^{20}\) cm\(^{-2}\). It is
therefore an extremely good candidate for studying the dynamical and radiative
processes which give rise to the X--ray emission in AGN.
In this letter we present the spectrum and spectral variability of \mrk\
as observed by the combination of instruments on board \xmm\ during the
performance verification (PV) observation. 

\section{The \xmm\ observation and data reduction}
\label{sec:observation}

The observation was performed on 20th May 2000
during \xmm\ revolution 0082 and lasted $\sim$60 ks. All instruments 
suffered some data
gaps close to the end of the observation.
The data were processed using the standard analysis software (SAS) pipeline.
For EPIC MOS1 and PN analysis we use the most up to date (October 2000) 
on-axis response matrices available. 
%MOS2 data are analysed using the MOS1
%response matrix as there is currently no MOS2 response matrix available. 
The RGS data were analysed using response matrices
generated by the SAS. 
%using the 12th July 2000 calibration release.

Due to high particle background near the beginning of the
observation, the EPIC PN and MOS cameras were closed for the
first 20 ks and 30 ks respectively. 
The EPIC MOS cameras were operated
in full frame imaging mode. The observed countrate of \mrk\ is about five times
larger than the recommended limit for this imaging mode, and therefore the MOS
data are piled up. This is expected to result in a hardening of the MOS
spectrum equivalent to a change in power law energy index of $\Delta \alpha
\sim 0.05$ (figure 28 in the \xmm\ User Handbook).
Apart from this 
spectral hardening, we do not expect photon pile-up to
seriously degrade the spectra. 
MOS spectra used in this paper were constructed using pattern 0-12 events.
Source data were taken from a forty arcsecond radius region around the source,
and  background data were taken from four identical circular regions close to 
the
source position. MOS Spectra were grouped to a minimum of 150 counts per group.

The EPIC PN camera was used in small window mode and is not significantly 
piled-up.
Source data were extracted from a 35 arcsecond radius region and
background data were taken from three identical circular regions. 
 Only single pixel (pattern 0)
events were used to construct PN spectra, 
to match the available response matrix.
In order to overcome a problem in current versions of the SAS, which produces 
periodic dips every $\sim$25 eV in PN spectra, 
we have manually grouped PN spectra to groups which are $>$ 25 eV. 
Less than 1\% of the counts in the source regions in the MOS and PN cameras 
are due to the background.

Both RGS cameras were operated in the standard spectroscopy+Q mode for
the duration of the observation. Source data were extracted using the standard
spatial and order selections, and background data were taken from two identical
sized extraction regions offset from the source in the cross dispersion 
direction. 87\% of the counts in the source regions come from
\mrk.

The OM took a sequence of 5 exposures
through the UVW1 filter, 10 exposures through the UVM2 filter, and 3
exposures through the UVW2 filter.

\section{The spectra}
\label{sec:spectra}

\begin{figure}
\begin{center}
\leavevmode
\setlength{\unitlength}{1cm}
\begin{picture}(8.8,12.4)
\put(0,0){\includegraphics{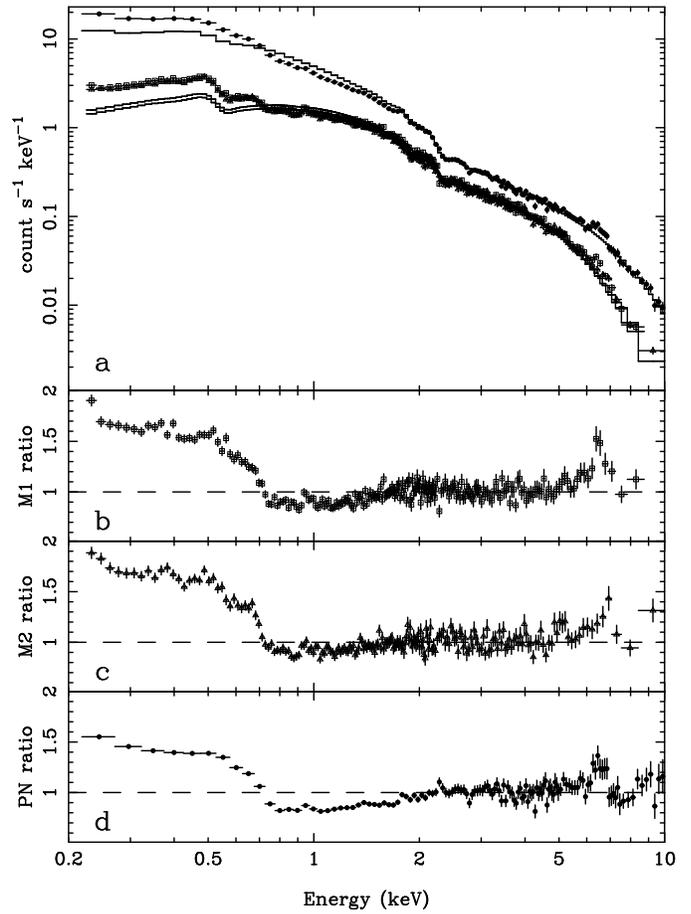}}
\end{picture}
\caption{a: spectra from MOS1 (open squares), MOS2 (open triangles),
 PN (filled dots) 
and power laws convolved with the
instrument response functions (stepped lines). b,c,d:
ratios of MOS1, MOS2, and PN spectra to the power law
model.}
\label{fig:epicspec}
\end{center}
\end{figure}

The EPIC PN, and MOS spectra of \mrk\ are shown in
Fig. \ref{fig:epicspec}a, and were analysed using the {\scriptsize XSPEC} 
software package. 

\subsection{2-10 keV spectrum}

\begin{table}
\caption{Power law ($F_{E} = K\ E^{-\alpha}$) 
fits to the EPIC spectra in the 2-5 and 7-10 keV energy 
ranges.}
\label{tab:plfits}
\begin{tabular}{lccc}
Instrument&$\alpha$&$K^{*}$&$\chi^{2}/\nu$\\
M1 &$0.98\pm0.03$&$4.66\pm0.16$&66.5/76\\
M2 &$0.96\pm0.03$&$4.30\pm0.16$&87.4/72\\
PN &$1.09\pm0.02$&$6.11\pm0.15$&71.8/71\\
PN high &$1.14\pm0.03$&$7.36\pm0.23$&12.7/18\\
PN low  &$1.05\pm0.03$&$5.21\pm0.18$&19.0/18\\
&&&\\
\multicolumn{4}{l}{$^{*}$ $10^{-3}$ keV cm$^{-2}$ s$^{-1}$ keV$^{-1}$}\\
\end{tabular}
\end{table}

We began studying the spectra by 
fitting a power law
model (\(F_{E} \propto E^{-\alpha}\)) with a Galactic \(\nh\) of 1.8
\(\times 10^{20}\) cm$^{-2}$ to the EPIC spectra in the energy ranges
2-5 and 7-10 keV where the spectrum
should be relatively unaffected by warm absorbers, Fe line emission or a
soft excess. 
Table \ref{tab:plfits} shows the results of these fits.
The data in these energy ranges are indeed fit well by a power
law model. The best fit slopes for the MOS data are harder than those for the 
PN data 
by $\Delta
\alpha \sim 0.1$; this is probably due to pile-up in the MOS 
cameras (see Section \ref{sec:observation}).
The data and model (including the 0.2-2 and 5-7 keV regions not used in the 
fit) are shown in Fig. \ref{fig:epicspec}a. The ratio of the
data to the model is shown for the three EPIC cameras in Fig. 
\ref{fig:epicspec}b,c, and d.

\begin{figure}
\begin{center}
\leavevmode
\setlength{\unitlength}{1cm}
\begin{picture}(8.8,5.2)
\put(0,0){\includegraphics{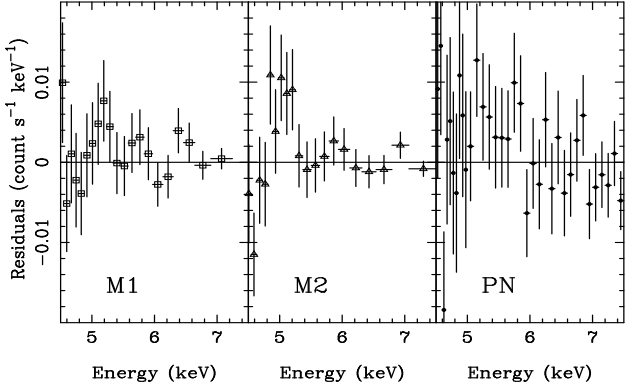}}
\end{picture}
\caption{Residuals of the gaussian fits to the Fe K$\alpha$ line (Table
\ref{tab:feline}). Note the excess emission at 5-5.5 keV indicating that the
emission line has a red wing.}
\label{fig:gauresids}
\end{center}
\end{figure}

A prominent feature, consistent with a broad Fe K$\alpha$ emission line 
centred at 6.4 keV, is present in the spectra from all 3 EPIC cameras.
An emission line component is strongly required by the data: fitting the whole
2-10 keV PN spectrum with only a power law and fixed Galactic \(\nh\) results
in a $\chi^{2}/\nu$ of 127/92, unacceptable at $>$ 99\% confidence.
The presence of Fe K$\alpha$ emission in \mrk\ has so far been controversial:
Leighy \etal (\cite{leighly96}) found evidence for a narrow Fe K$\alpha$ line,
Nandra \cite{nandra97} found evidence for a broad Fe K$\alpha$ line (both using
\asca\ data), while Matt
\etal \cite{matt00} found no strong evidence for an Fe K$\alpha$ line at all in
their \sax\ data. 
The results of fits to the 2-10 keV spectrum with
a power law continuum and Gaussian emission line model are given in Table
\ref{tab:feline}, and residuals to these fits are shown in
Fig. \ref{fig:gauresids}. 
The Fe K$\alpha$ line in Seyfert 1 galaxies is widely thought to arise from
fluorescence in near neutral material in an accretion disk around a black hole.
An emission line produced by this process typically has an asymmetric 
profile with significant broadening to the red.
The residuals to the gaussian line fits (Fig. \ref{fig:gauresids}) show an
excess of emission between 5 and 5.5 keV, suggesting that
the emission line in \mrk\ has such a red wing.
We have
therefore investigated accretion disk emission line models.
Table \ref{tab:disklinefits}, gives the results of model fits to the Fe K$\alpha$ 
line in the 2-10 keV PN data, which have the highest 
signal to noise. This time we have used emission
line profiles appropriate for accretion disks around a maximally rotating Kerr 
black hole (Laor \cite{laor91}, the {\scriptsize LAOR} model in 
{\scriptsize XSPEC}) and a Schwarzschild black hole (Fabian \etal 
\cite{fabian89}, the
{\scriptsize DISKLINE} model in {\scriptsize XSPEC}).
Both models fit the data well, 
and therefore the data are consistent with the Fe
K$\alpha$ line originating in an accretion disk. There is little to 
distinguish between the Kerr and Schwarzschild models
in terms of $\chi^{2}/\nu$.

\begin{table}
\caption{Fe K$\alpha$ line parameters when the 2-10 keV spectrum is fit with
a power law + Gaussian emission line.}
\label{tab:feline}
\begin{tabular}{lcccc}
Instrument&Energy$^{*}$&$\sigma$$^{*}$&EW$^{*}$&$\chi^{2}/\nu$\\
M1 &$6.5\pm0.2$        &$0.44^{+0.20}_{-0.16}$&0.38$^{+0.20}_{-0.08}$&78.2/88\\
M2 &$6.8^{+0.2}_{-0.5}$&$0.39^{+0.49}_{-0.21}$&0.26$^{+0.20}_{-0.16}$&110/83\\
PN &$6.5\pm0.1$        &$0.24^{+0.05}_{-0.05}$&0.19$^{+0.04}_{-0.03}$&90.4/89\\
&&&&\\
\multicolumn{5}{l}{$^{*}$ Rest frame in units of keV. EW is equivalent width.}\\
\end{tabular}
\end{table}

Matt et al. (2000) found that fits to their \sax\ spectra of \mrk\ were
significantly improved with the addition of an edge at 7.55 keV, or an ionized
reflection component (which produces an edge at a similar energy). We have 
repeated fits to our 2-10 keV spectra with power law and emission line 
models, adding either an edge at $\sim 7.5$ keV, or an ionized
reflector (the {\scriptsize PEXRIV} model in {\scriptsize XSPEC}).
For all three EPIC spectra, and for both additional components, 
improvements to the $\chi^{2}_{\nu}$ are less than 2$\sigma$ significant
according to the F-test, and we therefore conclude that the spectra do not 
require an ionized reflection component.
 
\begin{table}
\caption{Relativistic accretion disk fits to the Fe K$\alpha$ line in the 
PN data}
\label{tab:disklinefits}
\begin{tabular}{lll}
Model         &    Schwarzschild        &       Kerr                  \\
%&&\\
Energy$^{a}$  & $6.5\pm0.1$             & $6.6\pm0.1$                 \\
EW$^{a}$      & $0.29\pm0.07$           & $0.37\pm0.11$               \\
Inclination   & $34^{\circ}\pm7^{\circ}$& $30^{\circ}\pm6^{\circ}$    \\
$q^{b}$       & $2.5^{+1.6}_{-0.5}$     & $3.2^{+3.5}_{-0.4}$         \\
$\chi^{2}/\nu$& 86.3/88                 & 85.1/88                     \\
&&\\
\multicolumn{3}{l}{$^{a}$ Rest frame in units of keV. 
EW is equivalent width.}\\
\multicolumn{3}{l}{$^{b}$ power law dependence of emissivity on radius:}\\
\multicolumn{3}{l}{\ \ \ emissivity $\propto R^{-q}$. 
Inner disk radius fixed at 6 $R_{g}$}\\
\end{tabular}
\end{table}

\subsection{0.2-2 keV spectrum}

\begin{figure}
\begin{center}
\leavevmode
\setlength{\unitlength}{1cm}
\begin{picture}(8.8,8.8)
\put(0,0){\includegraphics{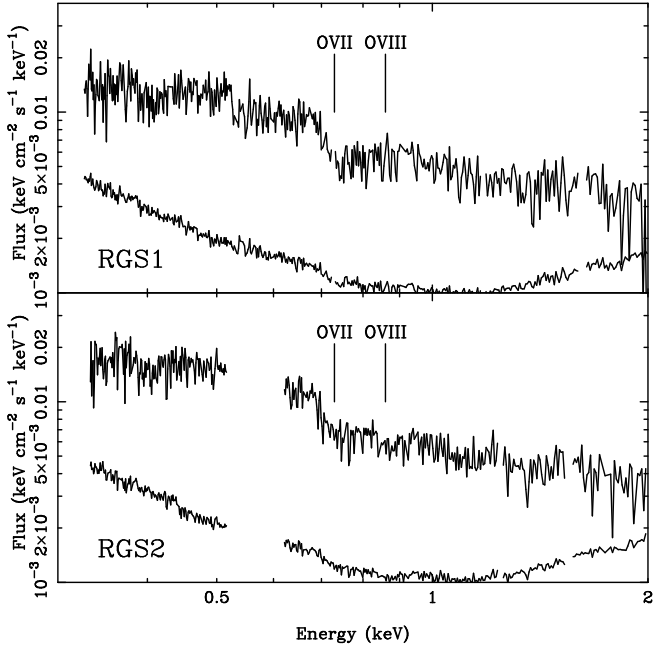}}
\end{picture}
\caption{Spectra of \mrk\ recorded by RGS1 (top panel) and RGS2 (bottom panel).
 The lower line in each panel shows the size of the 1 $\sigma$ statistical
uncertainty. The positions of the expected warm absorber OVII and OVIII 
K-edges are labelled.}
\label{fig:rgsspec}
\end{center}
\end{figure}

The EPIC data shown in Fig. 
\ref{fig:epicspec} show that between 
0.7 and 2 keV the spectrum is harder than the power law extrapolation. Below
 0.7 keV a strong excess is seen. 
Fig. \ref{fig:rgsspec} shows the spectrum of \mrk\ recorded by the two RGS, in
bins of 4 channels to improve signal to noise. The spectrum shows 
no strong narrow emission lines.
This rules out a significant nuclear starburst contribution to the soft X--ray
emission, which was suggested by Mathur (\cite{mathur00}). 
The most notable feature in the spectrum is a discontinuity at 0.7 keV. 
This feature corresponds to the
highest energy of the soft excess emission seen in the EPIC spectra. A
similar feature was
previously identified with the K-edge of OVII from a warm absorber 
in \asca\ and
\sax\ data by Leighly \etal (\cite{leighly96}) and Matt \etal (\cite{matt00})
respectively; the presence of an OVIII K-edge was also inferred from both these
datasets.  At the redshift of \mrk\ (0.0129) the OVII edge would be found at
0.730 keV and the OVIII edge at 0.860 keV; these energies are labelled in
Fig. \ref{fig:rgsspec}. Edges are not seen in the RGS data at either of these
energies. If the feature at 0.7 keV {\it is} an OVII K-edge, then the
absorbing material must be redshifted by more than 10000 km s$^{-1}$ which is
inconsistent with current physical scenarios for the warm absorber gas, eg that
it is in an outflowing wind (Kaastra \etal \cite{kaastra00}) or that it is
associated with the broad and/or narrow line regions (Otani \etal
\cite{otani96}). If the deficit of photons between 0.7 and 2.0 keV is not
produced by an ionized absorber, the continuum itself must deviate from a power
law shape, and have a harder spectrum in this energy range than at higher 
energies.

Detailed modelling of the soft X--ray spectrum of \mrk\ in this energy range 
is outside the scope of this letter. 
For a more detailed examination of 
the discrepancy between the
\mrk\ RGS data and warm absorber models, see Branduardi-Raymont
\etal (\cite{branduardi01}) who model the features in the
RGS spectrum as emission lines from a relativistic accretion disk.

\section{Lightcurves}
\label{sec:lightcurves}

\begin{figure}
\begin{center}
\leavevmode
\setlength{\unitlength}{1cm}
\begin{picture}(8.8,15)
\put(0,0){\includegraphics{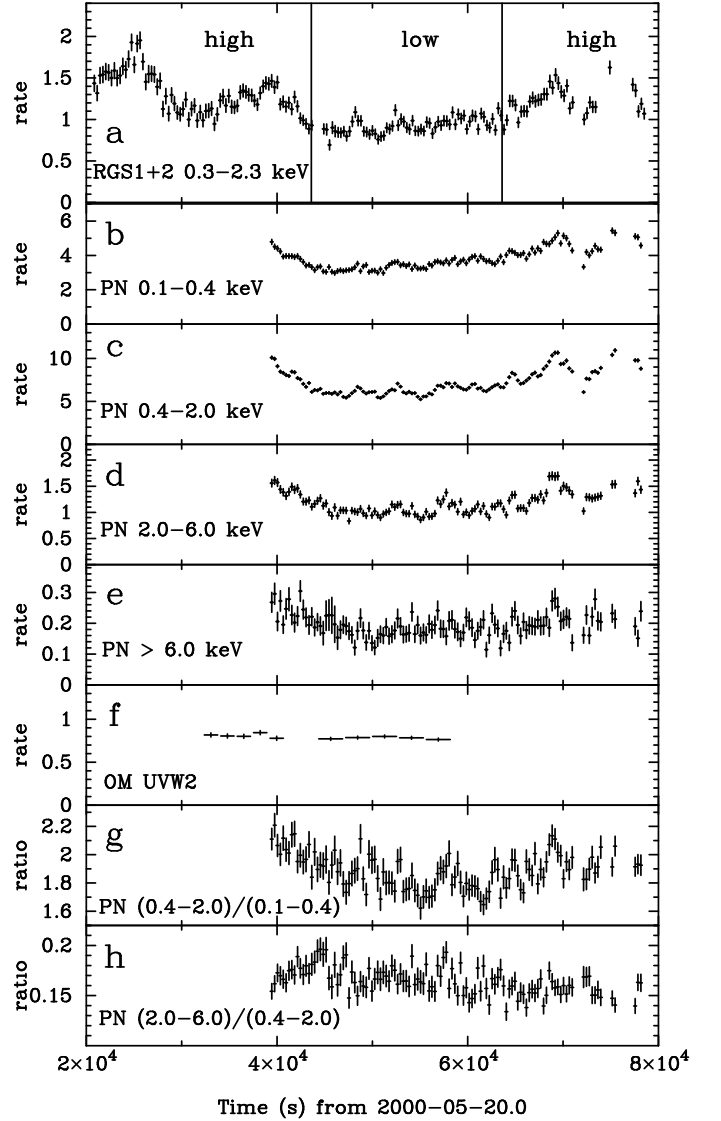}}
\end{picture}
\caption{Timeseries of \mrk\ in the RGS (a), EPIC PN (b,c,d,e) and OM through the UVM2 filter (f).  The y axis
units for panels a-f are count s$^{-1}$. Panels g and h show hardness ratios
constructed from the PN lightcurves, demonstrating that the source undergoes
significant spectral variability during the observation.}
\label{fig:lightcurves}
\end{center}
\end{figure}

RGS, EPIC PN and OM lightcurves of \mrk\ are shown in
Fig. \ref{fig:lightcurves}. The EPIC MOS lightcurves (not shown) are 
similar to the PN lightcurves but cover a shorter timespan and have lower 
signal to noise. The RGS and PN lightcurves are constructed with
300s bins. The PN lightcurves have been split into four different energy bands
(0.1-0.4 keV, 0.4-2.0 keV, 2.0-6.0 keV, and $>$ 6.0 keV) and are shown in 
panels b-e of Fig. \ref{fig:lightcurves}. 
Hardness ratios between the 0.1-0.4 keV,
0.4-2.0 keV and 2.0-6.0 keV energy bands are shown in panels g and h of
Fig. \ref{fig:lightcurves}. The low energy (0.4-2.0 keV)/(0.1-0.4 keV) hardness
ratio varies such that \mrk\ is harder when it is brighter. This is suggestive
of a scenario in which the variability of a broad band emission component is
diluted at the softest energies because of the superposition of a relatively
constant soft excess component. Although the correspondence is not exact, for
most of the observation the (2.0-6.0 keV)/(0.4-2.0 keV) hardness ratio appears
to show the opposite behaviour to the lower energy hardness ratio: the 0.4-6.0
keV spectrum is harder when the 0.1-2.0 keV spectrum is softer and vice versa.

We have quantified the variability in each of the PN lightcurves 
using the normalised 
variability amplitude (hereafter NVA)
which we define as:
%\begin{equation}
\[
{\rm NVA} = \sqrt{\frac{1}{\bar x (N-1)} \left( \left[
\sum^{N}_{i=1}  (x_{i}-\bar x)^{2} \right] -\sigma^{2} \right)}
\]
%\end{equation}
where $\sigma$ is the statistical 
uncertainty on the data points $x_{i}$.
1 $\sigma$ statistical uncertainties on the NVAs were estimated 
by bootstrap resampling of the
lightcurves. The NVAs of the 0.1-0.4 keV, 0.4-2.0 keV, 2.0-6.0 keV and $>$6.0
keV lightcurves are $0.15\pm0.01$, $0.19\pm0.01$, $0.17\pm0.01$ and
$0.12\pm0.02$ respectively. The 0.4-2.0 keV lightcurve shows more variability
than the low energy (0.4-2.0) and high
energy ($>$6.0 keV) lightcurves. 

The longest OM sequence of exposures (UVM2) is
shown in Fig. \ref{fig:lightcurves}f. 
In the ultraviolet there is little evidence for
variability: a constant is an acceptable fit to the UVM2 lightcurve 
(\(\chi^{2}/\nu\) of 11.1/9) 
and the data are all within 7\% of the mean
countrate. Similarly, no variability is found in the shorter UVW1 and
UVW2 sequences (not shown).

\section{Spectral variability}
\label{sec:spectralvariability}

\begin{figure}
\begin{center}
\leavevmode
\setlength{\unitlength}{1cm}
\begin{picture}(8.8,10.3)
\put(0,0){\includegraphics{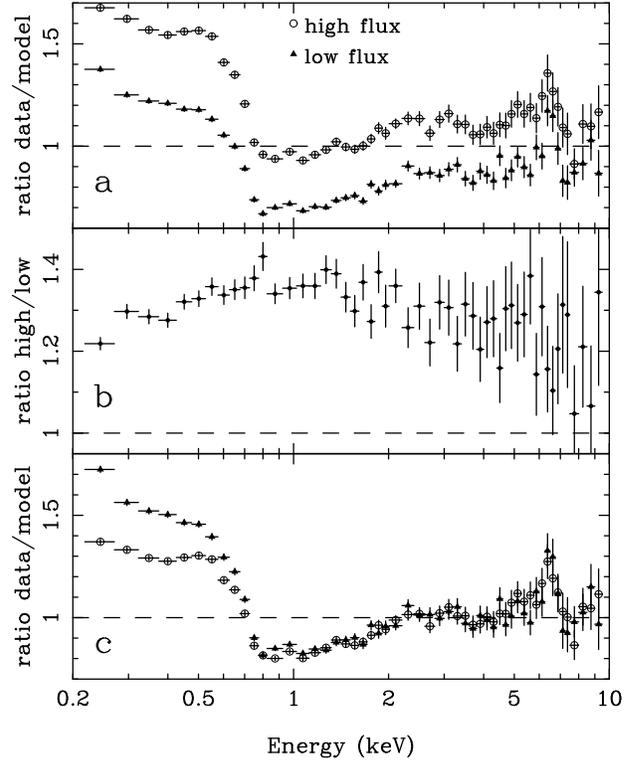}}
\end{picture}
\caption{a: ratio of high flux and low flux spectra 
(see Section \ref{sec:spectralvariability}) to the same power law model. The
open circles are used for the high flux spectrum and closed triangles for the
low flux spectrum.  b: ratio of the high and low flux spectra.  c: ratio of
high and low flux spectra to power law models fit individually to the 
two spectra, with symbols as in panel a.}
\label{fig:pn_ratio}
\end{center}
\end{figure}

To examine how the spectrum of \mrk\ changes with flux, we have
divided the observation into two parts corresponding to ``high flux'' and ``low
flux''. These time periods are indicated in
Fig. \ref{fig:lightcurves}a. The EPIC PN spectra from the high and low periods
are shown in 
Fig. \ref{fig:pn_ratio}a, after division by a model power law with energy index
$\alpha=1.09$ (the best fit slope for the 2-10 keV PN spectrum of the whole observation)
and 
fixed Galactic $\nh$.
In order to improve signal to noise, the PN high and low flux spectra have 
been grouped more heavily than the PN spectrum for the whole observation.
The main difference between the two spectra is that
the overall continuum is higher in the higher flux spectrum. 

The ratio
of the high to low flux spectra is shown in 
Fig. \ref{fig:pn_ratio}b. This shows that between 1 and 10
keV the high flux spectrum is softer than the low flux spectrum, while 
below 1 keV 
the high flux spectrum is harder than the low flux spectrum.
We have performed power law fits to the PN data in the 2-5 and 7-10 keV 
energy range for the high and low 
flux spectra separately, and the results of these fits are given in Table
\ref{tab:plfits}. The best-fit power law to the high flux spectrum is softer 
than the best fit to the low flux spectrum by $\Delta \alpha = 0.1$, 
significant at
2$\sigma$. 
Fig. \ref{fig:pn_ratio}c shows the ratio of the data to these best fit power
law models. The two ratios are indistinguishable except below 0.7 keV, where
the higher flux spectrum has a smaller fractional contribution from the soft 
excess emission. 
The overall spectral variability is therefore well characterised by an X--ray
continuum which softens as it increases in flux combined with a less variable 
or constant component at low energy. This scenario would naturally explain the
behaviour of the hardness ratios and
why the 0.4-2.0 keV lightcurve is the most variable of the PN lightcurves 
(Section \ref{sec:lightcurves}).
A variable warm absorber is not required to account for the spectral
variability (as was suggested by Matt \etal \cite{matt00}), because the 
deficit of photons between 0.7 and 2 keV relative to 
the extrapolated higher energy power law spectrum is the same in the high and
low flux spectra. 

\begin{figure}
\begin{center}
\leavevmode
\setlength{\unitlength}{1cm}
\begin{picture}(8.8,10.5)
\put(0,0){\includegraphics{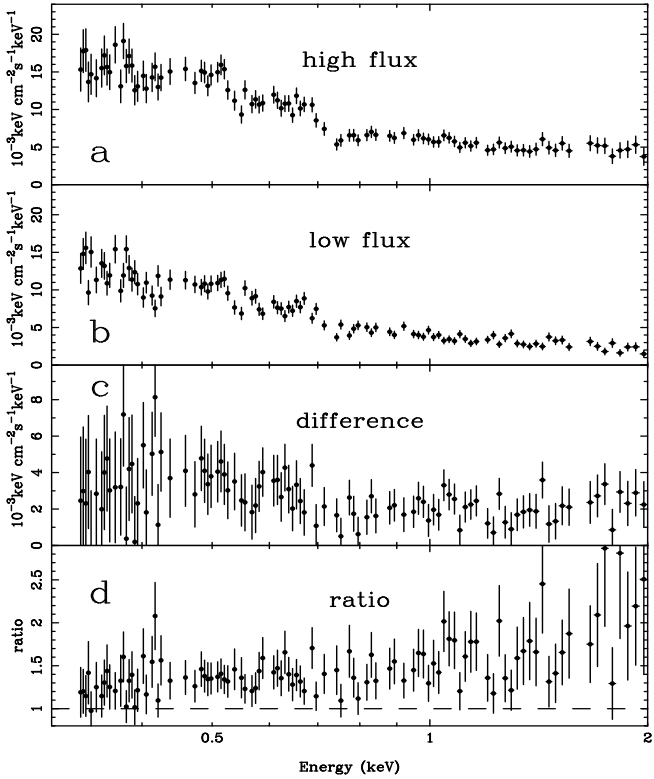}}
\end{picture}
\caption{a: high flux and b: low flux RGS1 spectra. c:
difference between the high and low flux spectra d: ratio of the high flux
spectrum to the low flux spectrum.}
\label{fig:rgshighlow}
\end{center}
\end{figure}

The spectral variability between 0.3 and 2 keV  
can be examined at higher resolution
using the RGS data. The high and low flux RGS1 spectra of \mrk, in 16 channel
bins, are shown in Fig. \ref{fig:rgshighlow}a,b. Fig. \ref{fig:rgshighlow}c 
shows the difference between these two spectra, and Fig. \ref{fig:rgshighlow}d
shows the ratio of the high flux to low flux spectra. 
As is found for the PN spectra, the ratio of the high flux to low flux RGS 
spectra is smaller close to 0.3 keV than
above 1 keV.
The 0.7 keV edge-like 
feature is present in the low flux spectrum and the high flux spectrum. It also
appears to be in the difference spectrum (though not the ratio spectrum) and
hence if this feature is produced by a relativistic emission
line, as proposed by Branduardi-Raymont \etal (\cite{branduardi01}), the line 
is varying with the continuum.

\section{Discussion}
\label{sec:discussion}

The standard paradigm for the medium energy X--ray continuum emission in 
radio-quiet AGN is that it is due to Compton up-scattering of soft UV
photons in some form of hot corona (Mushotzky, Done \& Pounds
\cite{mushotzky93} and references therein). 
The seed UV photons could be
produced either by an accretion disk (Haardt, Maraschi \& Ghisellini
\cite{haardt97}) or
from low temperature (T $\sim 3 \times 10^{5}$ K) thermal emission in the
corona itself (Torricelli-Ciamponi \& Courvoisier \cite{torricelli95}).  In
\mrk\ the broad Fe K$\alpha$ line seen in the EPIC spectra provides strong
evidence for the presence of a cool, optically thick accretion disk (Tanaka
\etal \cite{tanaka95}). This, and
the lack of narrow coronal emission lines in the long wavelength part of the
RGS spectra, favours the accretion disk as the source of the UV-soft X--ray
bump in \mrk. 

%In the two phase disk-corona model of Haardt, \etal (\cite{haardt97}) the hot
%corona is Compton cooled by the soft photons from the disk, while the disk is
%heated by irradiation from the corona. 
%This feedback mechanism results in a 
%strong coupling between the temperature and optical depth of the corona and the
%soft excess emission from the disk. 
%If the corona is sufficiently compact, and luminous at high energy ($>$ 1 MeV),
%the plasma may become pair dominated.  
%In this case, a noticeable change in spectral index of the medium energy
%spectrum is only expected for large amplitude ($>$ factor 3) changes 
%in the 2-10 keV luminosity (Haardt \etal
%\cite{haardt97}). The change in spectral index with moderate (factor 2)
%variability seen during the \xmm\ observation is therefore consistent with a
%low pair density corona in \mrk.

The UV - soft X--ray accretion disk emission is effectively bracketed by 
the soft X--ray excess at $^{<}_{\sim}0.7$ keV, and the UV emission observed in
the OM.
The behaviour of the
(0.4-2.0 keV)/(0.1-0.4 keV) hardness ratio (Section \ref{sec:lightcurves}), and
the X--ray spectral variability (Section \ref{sec:spectralvariability}) 
implies that
the soft excess varies significantly less than the Comptonised continuum.
Furthermore, the OM lightcurves are consistent with zero variability in the 
UV.  
It therefore appears that the Compton upscattered X--ray continuum varies much
more than its source of seed photons, the UV - soft X--ray bump.
%It therefore appears that the whole UV - soft X--ray bump varies much less than
%the Compton upscattered X--ray continuum during the observation.  
This implies that either the
Comptonising corona is highly susceptible to changes in the soft photon
field or that the variability of the Comptonised component is intrinsic to 
the corona itself and is unrelated to variations in the soft photon field.
Something specific to the corona such as changing magnetic fields or
the spectrum of particles injected into the corona would therefore provide a 
good explanation for the X--ray variability.

\section{Conclusions}

We have presented RGS, EPIC and OM spectra and lightcurves of \mrk\ from the 60
ks \xmm\ PV phase observation. In all three EPIC cameras the spectra above 2
keV are well described by a power law 
with a broad Fe K$\alpha$ emission line. Relative to the power law, the EPIC
spectra show a deficit of photons between $\sim$ 0.7 and 2
keV and excess emission at lower energies.  The RGS spectra contain no strong,
narrow
emission lines. Warm absorber edges are not seen at the expected energies of
OVII and OVIII in the RGS spectra; 
an edge-like feature at 0.7 keV could only be due to OVII
if the absorbing material were redshifted more than 
10,000 km/s relative to the system. 
Significant variability is detected in the X--ray lightcurves and hardness
ratios, while no variability is detected in the UV with the OM.
A variable warm absorber is not required to account for the spectral
variability, which can be explained by a medium energy continuum component
which softens as it increases in flux, with an additional contribution from 
a less variable emission component at low energy. 
%Two phase disk-corona 
%feedback models are compatible with 
%this if the corona has a low pair density.
Changes in magnetic fields threading the accretion disk corona, or the 
spectrum
of particles injected into the corona, can explain the observed spectral 
variability.

\end{document}